\documentclass[10pt, conference, letterpaper]{IEEEtran}

\usepackage{cite}
\usepackage{enumerate}
\usepackage{graphicx}
\usepackage[cmex10]{amsmath} 
\usepackage{amssymb}
\usepackage{xspace}
\usepackage{subfigure}
\usepackage[lined,linesnumbered,ruled,commentsnumbered]{algorithm2e} 
\usepackage{colonequals}
\usepackage[mathscr]{euscript}
\usepackage{fixltx2e}    
\usepackage{amsthm}
\usepackage{mathrsfs}

\usepackage{xspace}

\usepackage{epsfig}
\usepackage{epstopdf}

\usepackage{tikz}
\usepackage{pgfplots}

\usetikzlibrary{arrows,shapes,backgrounds,plotmarks,positioning}

\pgfplotsset{compat=newest}                         
\pgfplotsset{plot coordinates/math parser=false}
\newlength\figureheight
\newlength\figurewidth

\newtheorem{theorem}{Theorem}
\newtheorem{lemma}{Lemma}
\newtheorem{definition}{Definition}

\newtheorem{example}{Example}
\newtheorem{remark}{Remark}
\newtheorem{experiment}{Experiment}

\newcommand{\RZ}[1]{\mathsf{Z}_{#1}}

\newcommand{\RW}[1]{\mathsf{W}_{#1}}

\newcommand{\RRCO}{\mathscr{R}}

\newcommand{\Card}[1]{|#1|}
\newcommand{\Set}[1]{\{#1\}}
\newcommand{\Core}[1]{\mathscr{C}(#1)}

\newcommand{\G}[2]{\Omega(#1,#2)}

\newcommand{\Pat}{\mathcal{P}}

\newcommand{\rv}{\mathbf{r}} 
\newcommand{\rvS}{\hat{\mathbf{r}}} 
\newcommand{\rS}{\hat{r}} 

\newcommand{\Real}{\mathbb{R}}
\newcommand{\RealP}{\mathbb{R}_{+}}    

\newcommand{\EX}{\text{EX}}

\newcommand{\Hdual}{H^{\#}}

 {\begin{list}{}%
         {\setlength{\leftmargin}{#1}}%
         \item[]%
 }
 {\end{list}}

\begin{document}

\title{Fairness in Multiterminal Data Compression: Decomposition of Shapley Value}

\author{\IEEEauthorblockN{Ni~Ding\IEEEauthorrefmark{1},
                          David~Smith\IEEEauthorrefmark{1},
                          Parastoo~Sadeghi\IEEEauthorrefmark{2} and
                          Thierry~Rakotoarivelo\IEEEauthorrefmark{1}}
        \IEEEauthorblockA{\IEEEauthorrefmark{1}Data61, The Commonwealth Scientific and Industrial Research Organisation
            \\$\{$ni.ding, david.smith, thierry.rakotoarivelo$\}$@data61.csiro.au}
        \IEEEauthorblockA{\IEEEauthorrefmark{2}The Research School of Engineering, College of Engineering and Computer Science, The Australian National University
            \\$\{$parastoo.sadeghi$\}$@anu.edu.au}
    }

%


\maketitle

\begin{abstract}
We consider the problem of how to determine a fair source coding rate allocation method for the lossless data compression problem in multiterminal networks, e.g, the wireless sensor network where there are a large number of sources to be encoded. We model this problem by a game-theoretic approach and present a decomposition method for obtaining the Shapley value, a fair source coding rate vector in the Slepian-Wolf achievable region. We formulate a coalitional game model where the entropy function quantifies the cost incurred due to the source coding rates in each coalition. In the typical case for which the game is decomposable, we show that the Shapley value can be obtained separately for each subgame. The complexity of this decomposition method is determined by the maximum size of subgames, which is strictly smaller than the total number of sources and contributes to a considerable reduction in computational complexity. Experiments demonstrate large complexity reduction when the number of sources becomes large.
\end{abstract}


\section{introduction}

In wireless communications, we usually discuss how to transmit the messages in the most efficient way and many studies deal with the problem of collecting and forwarding information from a random source to the sink or destination with the minimum cost, e.g., \cite{Xiong2004,Roua2010}. For multiple random sources, which are correlated in general, using the minimal code length to describe the sources with the least information loss is referred to as the \textit{data compression}, or source coding, problem \cite{Cover2012ITBook}.
While the authors in \cite{SW1973,Cover1975} derived the Slepian-Wolf (SW) constraints (lower bounds) that describe the achievable source coding rate region, they also stated that the lossless data compression can be attained without interactive communications between the sources, i.e., the individual sources can be encoded in a lossless manner without the knowledge of others. This allows the information theoretic results on source coding problems to be applied to multiterminal networks, e.g., peer-to-peer (P2P) networks.

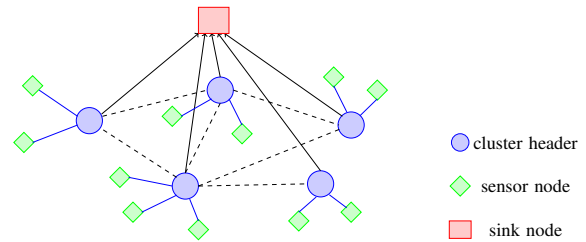
\begin{figure}[tpb]
	\centering
    \scalebox{0.58}{\begin{tikzpicture}

\draw [color = blue, fill = blue!20] (0,0.5) circle (0.3);
\draw [->] (0.28,0.65)--(2.5,2.5);
\draw [color = blue, fill = blue!20] (3,1.2) circle (0.3);
\draw [->] (3,1.5)--(2.8,2.5);
\draw [color = blue, fill = blue!20] (2.2,-1) circle (0.3);
\draw [->] (2.2,-0.7)--(2.65,2.5);
\draw [color = blue, fill = blue!20] (5.3,-0.95) circle (0.3);
\draw [->] (5.3,-0.65)--(2.95,2.5);
\draw [color = blue, fill = blue!20] (6,0.4) circle (0.3);
\draw [->] (5.8,0.6)--(3.1,2.5);

\draw [color = red, fill = red!20] (2.5,2.5) rectangle (3.2,3.1);

\node [draw, diamond, color = green, fill = green!20] at (-1.3,1.3) {};
\node [draw, diamond, color = green, fill = green!20] at (-1.5,0)  {};
\draw [color = blue] (-0.27,0.4)--(-1.3,0);
\draw [color = blue] (-0.27,0.6)--(-1.15,1.2);

\node [draw, diamond, color = green, fill = green!20] at (0.7,-0.8) {};
\node [draw, diamond, color = green, fill = green!20] at (1,-1.6) {};
\node [draw, diamond, color = green, fill = green!20] at (2.5,-2) {};
\draw [color = blue] (1.9,-1)--(0.9,-0.8);
\draw [color = blue] (2,-1.21)--(1.2,-1.6);
\draw [color = blue] (2.3,-1.3)--(2.5,-1.8);

\node [draw, diamond, color = green, fill = green!20] at (1.9,0.6) {};
\node [draw, diamond, color = green, fill = green!20] at (3.5,0.2) {};
\draw [color = blue] (2.1,0.6)--(2.8,1);
\draw [color = blue] (3.2,1)--(3.5,0.4);

\node [draw, diamond, color = green, fill = green!20] at (5.6,1.5) {};
\node [draw, diamond, color = green, fill = green!20] at (6.6,1.2) {};
\draw [color = blue] (5.9,0.7)--(5.6,1.3);
\draw [color = blue] (6.2,0.61)--(6.6,1);

\node [draw, diamond, color = green, fill = green!20] at (4.8,-1.8) {};
\node [draw, diamond, color = green, fill = green!20] at (6,-1.6) {};
\draw [color = blue] (5.2,-1.25)--(4.8,-1.6);
\draw [color = blue] (5.5,-1.2)--(5.9,-1.5);

\draw [dashed] (2,-0.8)--(0.3,0.3);
\draw [dashed] (2.2,-0.7)--(3,0.9);
\draw [dashed] (3.3,1.2)--(5.72,0.4);
\draw [dashed] (0.3,0.6)--(2.7,1.2);
\draw [dashed] (2.5,-1)--(5.73,0.3);
\draw [dashed] (2.5,-1)--(5,-0.95);

\draw [color = blue, fill = blue!20] (8.5,0) circle (0.2);
\node at (10,0) {\large cluster header};
\node [draw, diamond, color = green, fill = green!20] at (8.5,-1) {};
\node at (10,-1) {\large sensor node};
\draw [color = red, fill = red!20] (8.25,-2.2) rectangle (8.75,-1.8);
\node at (10,-2) {\large sink node};

\end{tikzpicture}}
	\caption{Wireless sensor network: A group of sensor nodes measure some features in a region. Sensors form clusters with one in each cluster serving as the cluster header to forward the measurements to a sink node. The cluster headers may or may not be connected. }
	\label{fig:WSN}
\end{figure}

The typical examples of the multiterminal source coding problem are the coded cooperative data exchange (CCDE) problem \cite{Roua2010,SprintRand2010}, where the sources are mobile clients, and the distributed source coding problem over a wireless sensor network (WSN) \cite{Xiong2004,Srisooksai2012}. Take a WSN for example. We usually have a cluster, or group, of sensor nodes scattered in one area and one of them is served as the cluster header to collect the measurements and forward to a sink node or base station. See Fig.~\ref{fig:WSN}. Such a WSN poses a two-layer multiterminal source coding problem: sensors-to-cluster header and cluster headers-to-base station, which is usually solved in a distributed manner, e.g., \cite{Schonberg2004}.

Multiterminal data compression usually refers to the case of more than two data sources. In some cases, such as for a WSN, there are a large number of sources to be encoded. Importantly, it was pointed out in \cite{Madiman2008ISIT,Madiman2008} that all source coding rates for the lossless multiterminal data compression constitute the core of a coalitional game and the core can be very large.
For the problem of how to select a solution in the core, the fairness is usually considered in multiterminal networks such that the users are equally privileged, e.g., CCDE and WSN.\footnote{A fair source coding scheme also evens out the battery power consumption of sensors, which prolongs the lifetime of a WSN: The lifetime is usually defined as the time to which the first sensor node runs out of battery power\cite{Dietrich2009}.}
While the fair source coding schemes for the two terminals case were proposed in \cite{Pradhan2000,Sartipi2005} for a WSN, the question remains as to how to allocate the source coding rates fairly when the number of terminals becomes large. The authors in \cite{Madiman2008} proposed a fairness solution by the Shapley value. But, the exponentially
growing complexity of obtaining such a value imposes huge computation burden in large scale systems.

This paper addresses the problem of fairness in the multiterminal data compression problem from a coalitional game-theoretic point of view. We propose a decomposition method for obtaining the Shapley value \cite{Shapley1953Value}, a fair source coding rate vector in the core.
We first convert the SW lower bounds to upper bounds, based on which a coalitional game model is formulated with the entropy function quantifying the cost incurred due to the source coding rates in each coalition.
The submodularity of the entropy function ensures the convexity of the game and also ensures that the core, or the achievable rate vector set for the lossless data compression, is nonempty. We then explain the incentive for the users to cooperate: The users pay the least cost only if they cooperatively encode the multiple sources.
We show that, in the common case that the game can be decomposed into a group of subgames, the Shapley value can be obtained separately for each subgame. While the finest decomposer of a decomposable game can be determined by an existing algorithm in \cite{Bilxby1985} in quadratic time, the complexity of obtaining the Shapley value by the decomposition method is determined by the maximal subgame size, which is strictly less than the total number of sources in the entire system and contributes to a considerable reduction in computational complexity. Experiments demonstrate large reduction when the number of sources becomes large.

\subsection{Organization}
The rest of the paper is organized as follows. In Section~\ref{sec:system}, we describe the system model for the multiterminal source coding problem. A coalition game model characterized by the entropy function of the random sources is proposed in Section~\ref{sec:system}, where we show the convexity of the game and explain the users' incentive to cooperate in the game. In Section~\ref{sec:Shapley}, we propose the decomposition method to obtain the Shapley value and discuss the method's complexity. In Section~\ref{sec:Conclusion}, we provide some concluding remarks.

\section{System Model}
\label{sec:system}

Let $V$ with $\Card{V}>1$ be a finite set that contains the indices of all users in the system. We call $V$ the \textit{ground set}. Let $\RZ{V}=(\RZ{i}:i\in V)$ be a vector of discrete random variables indexed by $V$. User $i$ privately observes an $n$-sequence $\RZ{i}^n$ of the random source $\RZ{i}$ that is i.i.d.\ generated according to the joint distribution $P_{\RZ{V}}$. The users are required to send their observations to a sink node $T$ in a way so that the source sequence $\RZ{V}^{n}$ can be reconstructed by $T$. This problem is called \textit{multiterminal data compression} or \textit{source coding with side information} \cite{Cover2012ITBook}.

Let $\rv_V = (r_i \colon i \in V)$ be a rate vector that is indexed by $V$. Each dimension $r_i$ in $\rv_V$ denotes the source coding rate of $\RZ{i}$, i.e., the expected coding length at which user $i$ encodes his/her observation $\RZ{i}^{n}$ of source $\RZ{i}$.
We call $\rv_V$ an \textit{achievable rate vector} if the sink $T$ is able to recover the source sequence $\RZ{V}^{n}$ by letting the users transmit at the rates that are designated by $\rv_V$. The achievable rate region for the lossless data compression is characterized by the Slepian-Wolf (SW) constraints as follows.

\subsection{Achievable Rate Region: Slepian-Wolf Constraints}
\label{sec:SW}

For $X,Y \subseteq V$, let $H(\RZ{X})$ be the amount of randomness in $\RZ{X}$ measured by Shannon entropy \cite{Cover2012ITBook}\footnote{In this paper, we take all logarithms to base 2 so that all entropies are measured in bits. However, the results in this paper apply to all bases.} and $H(\RZ{X}|\RZ{Y})=H(\RZ{X \cup Y})-H(\RZ{Y})$ be the conditional entropy of $\RZ{X}$ given $\RZ{Y}$. In the rest of this paper, without loss of generality, we simplify the notation $\RZ{X}$ by $X$.
For a rate vector $\rv_V$, let $r$ be the \textit{sum-rate function} associated with $\rv_V$ such that
$$ r(X)=\sum_{i\in X} r_i, \quad \forall X \subseteq V $$
with the convention $r(\emptyset)=0$. $r(X)$ is the sum of source coding rates over all the users in $X$ and $r(V)$ is the overall source coding rate in the system.

It is shown in \cite{SW1973,Cover1975} that the sources can be encoded separately at rates $\rv_V$ so that the sink node $T$ is able to decode them with arbitrarily small error probability if and only if\footnote{In this paper, we consider the perfect case of data compression $r(V) = H(V)$ when there is no information redundancy, which allows us to derive the decomposition property in Section~\ref{sec:Decomposition}. However, one can allow $r(V) \geq H(V)$, or $r(C) \geq H(C)$ for each subgame if the game is decomposable (see Section~\ref{sec:Decomposition}), in the implementation.}
\begin{equation} \label{eq:SWConstrs}
    \begin{aligned}
        r(X) & \geq H(X|V \setminus X), \quad \forall X \subsetneq V, \\
        r(V) & = H(V).
    \end{aligned}
\end{equation}
The interpretation of \eqref{eq:SWConstrs} is: For the data compression to be lossless, (a) the users in $X$ must reveal $H(X|V \setminus X)$, the information that is uniquely obtained by $X$, to the sink; (b) the users must reveal the total information $H(V)$ to the sink. The inequalities in \eqref{eq:SWConstrs} are called the SW constraints, which describe the achievable rate region
\begin{multline}
    \RRCO(V) = \Set{ \rv_V \in \Real^{|V|} \colon  \\     r(X) \geq H(X|V \setminus X), \forall X \subsetneq V,
                        r(V) = H(V) }.  \nonumber
\end{multline}
In Section~\ref{sec:BasePoly}, we show that $\RRCO(V) \neq \emptyset$ due to the submodularity of the entropy function \cite{FujishigePolyEntropy}. In Section~\ref{sec:Intepretation}, this nonemptiness of $\RRCO(V)$ will be interpreted from a coalitional game-theoretic point of view.

\begin{figure}[tpb]
	\centering
    \scalebox{1}{\begin{tikzpicture}

\draw (-2,0) circle (0.5);
\node at (-2,0) {\Large $1$};
\node at (-2.3,-0.7) {\scriptsize \textcolor{blue}{$\RZ{1} = (\RW{a},\RW{b},\RW{c},\RW{d},\RW{e})$}};

\draw (2,0) circle (0.5);
\node at (2,0) {\Large $2$};
\node at (2,-0.7) {\scriptsize \textcolor{blue}{$\RZ{2} = (\RW{a},\RW{b},\RW{f})$}};

\draw (0,-1) circle (0.5);
\node at (0,-1){\Large $3$};
\node at (0,-1.7) {\scriptsize \textcolor{blue}{$\RZ{3} = (\RW{c},\RW{d},\RW{f})$}};

\draw (0,1.5) circle (0.5);
\node at (0,1.5){\Large $T$};

\draw [->] (-1.7,0.4)--(-0.4,1.2);
\draw [->] (1.7,0.4)--(0.4,1.2);
\draw [->] (0,-0.5)--(0,1);

\end{tikzpicture} }
	\caption{The multiterminal data compression problem with $V = \Set{1,2,3}$ in Example~\ref{ex:main}: User $i$ observes $\RZ{i}$ in private. The users are required to encode their sources $\RZ{i}$s so that the sink node $T$ is able to recover $\RZ{V}$. }
	\label{fig:SC}
\end{figure}
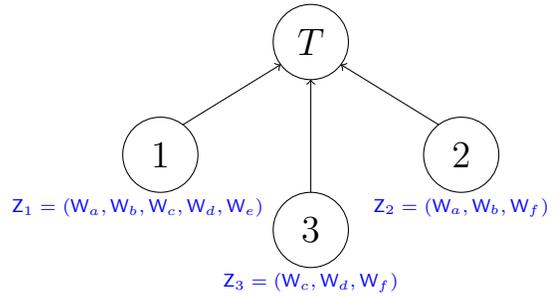

\begin{example} \label{ex:main}
Assume there are three users $V=\Set{1,2,3}$ in the system. They observe respectively
    \begin{align}
        \RZ{1} &= (\RW{a},\RW{b},\RW{c},\RW{d},\RW{e}),   \nonumber\\
        \RZ{2} &= (\RW{a},\RW{b},\RW{f}),   \nonumber\\
        \RZ{3} &= (\RW{c},\RW{d},\RW{f}),   \nonumber
    \end{align}
where $\RW{b}$ is an independent random bit with $H(\RW{b}) = \frac{3}{10}$, $\RW{f}$ is an independent random bit with $H(\RW{f}) = \frac{1}{2}$ and all other $\RW{j}$ are independent uniformly distributed random bit, i.e., $H(\RW{j}) = 1$ for all $j \in \Set{a,c,d,e}$. The users are required to encode $\RZ{i}$s so that $\RZ{V}$ can be reconstructed at $T$. See Fig.~\ref{fig:SC}.

In this system, the achievable rate region is characterized by the SW constraints:
    \begin{equation}
        \begin{aligned}
            \RRCO(V)=\big\{ \rv_V\in\Real^{|V|} \colon & r(\emptyset)=0,  \\
            &r(\Set{1}) \geq H(\Set{1}|\Set{2,3})=1 ,   \\
            &r(\Set{2}) \geq H(\Set{2}|\Set{1,3})=0 ,   \\
            &r(\Set{3}) \geq H(\Set{3}|\Set{1,2})=0 ,   \\
            &r(\Set{1,2}) \geq H(\Set{1,2}|\Set{3})=\frac{23}{10} ,  \\
            &r(\Set{1,3}) \geq H(\Set{1,3}|\Set{2})=3 ,   \\
            &r(\Set{2,3}) \geq H(\Set{2,3}|\Set{1})=\frac{1}{2} \big\} \\
            &r(\Set{1,2,3}) = \frac{24}{5}.
        \end{aligned} \nonumber
    \end{equation}
We have $\RRCO(V) \neq \emptyset$. For example, $\rv_V = (1, \frac{9}{5}, 2)$ is an achievable rate vector in $\RRCO(V)$.
\end{example}

\section{Coalitional Game}
\label{sec:game}

The relationship between the data compression problem and the coalitional game formulation was previously revealed in \cite{Madiman2008ISIT,Madiman2008} based on the SW constraints \eqref{eq:SWConstrs}. It is shown that the achievable rate region $\RRCO(V)$ coincides with the core of a convex game which is necessarily nonempty. In this section, we express the achievable rate region $\RRCO(V)$ by converting the lower bounds in \eqref{eq:SWConstrs} to the upper bounds in terms of the entropy function, based on which, we propose a coalitional game model. We show the equivalence of the core and the submodular base polyhedron of the entropy function, which allows us to derive the decomposition property of the core in Section~\ref{sec:Shapley} and address  a vital aspect missing from \cite{Madiman2008ISIT,Madiman2008}: The reason for the users to be cooperative in the multiterminal data compression problem in $V$.

\subsection{Base Polyhedron}
\label{sec:BasePoly}

For $X \subseteq V$, consider the lower bound $r(X) \geq H(X | V \setminus X)$ in the SW constraints~\eqref{eq:SWConstrs}. Since we also restrict the total rate $r(V) = H(V)$, we necessarily impose an upper bound constraint on the sum-rate in $V \setminus X$
$$r(V \setminus X)=r(V)-r(X) \leq H(V) - H(X | V \setminus X) = H(V \setminus X). $$
By converting the lower bounds in \eqref{eq:SWConstrs} for all $X \subseteq V$, we have the SW constrained region fully characterized by the entropy function $H \colon 2^V \mapsto \Real$:
\begin{equation} \label{eq:SWConstrsDual}
    \begin{aligned}
        r(X) & \leq H(X), \quad \forall X \subsetneq V, \\
        r(V) & = H(V).
    \end{aligned}
\end{equation}

It is shown in \cite{FujishigePolyEntropy} that the entropy function $H$ is a \textit{polymatroid rank function}: (a) normalized: $H(\emptyset) = 0 $; (b) monotonic: $H(X) \geq H(Y)$ for all $X,Y \subseteq V$ such that $Y \subseteq X$; (c) \textit{submodular}:
$$ H(X) + H(Y) \geq H(X \cap Y) + H(X \cup Y), \quad X, Y \subseteq V.$$
The polyhedron and base polyhedron of $H$ are respectively \cite[Section 2.3]{Fujishige2005} \cite[Definition 9.7.1]{Narayanan1997Book}
\begin{equation}
    \begin{aligned}
        & P(H,\leq) = \Set{\rv_V\in\Real^{|V|} \colon r(X) \leq H(X),\forall X \subseteq V},  \\
        & B(H,\leq) = \Set{\rv_V \in P(f,\leq) \colon r(V) = H(V)}.  \\
    \end{aligned}  \nonumber
\end{equation}
We have the achievable rate region coincides with the base polyhedron of the entropy function $H$:
$$ \RRCO(V) = B(H,\leq).$$

\subsection{Coalitional Game Model}

Let the users in $V$ be the players in a game. Instead of being totally selfish, the players may cooperate with each other to form coalitions, e.g., instead of encoding the source $\RZ{i}$ by him/herself, user $i$ may form the group $X$ with the other users in $X \setminus \Set{i}$ so as to encode $\RZ{X}$ cooperatively.
In this game, a subset $X \subseteq V$ is called a \textit{coalition} and $V$ is the \textit{grand coalition}. Let $H$ be the \textit{characteristic cost} function such that, for each coalition $X \subseteq V$, $H(X)$ quantifies the outcome of coalition $X$: If all the players $i \in X$ cooperate in $X$, i.e., if coalition $X$ forms, the total cost incurred due to the source coding rates among the users in $X$ is $H(X)$.
The coalitional game model for the multiterminal data compression problem is characterized by the ground set $V$ that indexes all the players and the entropy function $H$. We denote the coalition game model by $\G{V}{H}$.

In $\G{V}{H}$, $\rv_V$ is a cost allocation method that describes how the total cost $r(V)$ is distributed to the individual users. Here, the source coding rate $r_i$ is interpreted as the cost that user $i$ pays in participating the data compression problem in $V$.
The solution of the game $\G{V}{H}$ is the \textit{core} \cite{Shoham2008,Shapley1969Core}
\begin{multline} \label{eq:Core}
    \Core{V} = \{ \rv_V\in\Real^{|V|} \colon r(X) \leq H(X),\forall X \subseteq V, \\ r(V)=H(V) \},
\end{multline}
which contains all cost allocation methods $\rv_V$ that distribute exactly cost $H(V)$ among the users in $V$. The inequality $r(X) \leq H(X)$ states that the total the cost in any coalition $X \subsetneq V$ is no greater than $H(X)$, which ensures that no users have incentive to break from the grand coalition and form a smaller one \cite{Shoham2008}. In this sense, the core contains all cost/rate allocation methods such that all users would like to take part in the data compression problem in $V$.\footnote{It is also show in \cite[Theorem 8]{Shapley1971Convex} that the core is a stable set of cost allocation methods such that no coalitions $X \subsetneq V$ have incentive to break from the grand coalition.}
Based on the definition of the core \eqref{eq:Core}, it is easy to see the equivalence of the achievable rate region, the base polyhedron of $H$ and the core:
\begin{equation} \label{eq:Equivalence}
    \RRCO(V) = B(H,\leq) = \Core{V}.
\end{equation}
This equivalence also gives rise to the main result in this paper: The decomposition of $B(H,\leq)$ is equivalent to the decomposition of the game $\G{V}{H}$, which results in the decomposition of the Shapley value (See Section~\ref{sec:Decomposition}).

\begin{figure}[tbp]
	\centering
    \scalebox{0.7}{
%
%
%
\definecolor{mycolor1}{rgb}{0.5,0.5,0.9}%
\begin{tikzpicture}

\begin{axis}[%
width=3in,
height=2.5in,
view={-33}{30},
scale only axis,
xmin=0,
xmax=4.5,
xlabel={\Large $r_1$},
xmajorgrids,
ymin=0,
ymax=2.5,
ylabel={\Large $r_2$},
ymajorgrids,
zmin=0,
zmax=2.8,
zlabel={\Large $r_3$},
zmajorgrids,
axis x line*=bottom,
axis y line*=left,
axis z line*=left,
legend style={at={(0.5,0.9)},anchor=north west,draw=black,fill=white,legend cell align=left}
]

\addplot3[area legend,solid,fill=mycolor1,draw=black]
table[row sep=crcr]{
x y z\\
4.3 0.5 0 \\
4.3 0 0.5 \\
2.3 0 2.5 \\
1 1.3 2.5 \\
1 1.8 2 \\
3 1.8 0 \\
};
\addlegendentry{\Large $\RRCO(V) = B(H,\leq) = \Core{V}$};

\addplot3[area legend,solid,fill=white!90!black,opacity=4.000000e-01,draw=black]
table[row sep=crcr]{
x y z\\
4.3 0 0 \\
4.3 0.5 0 \\
4.3 0 0.5 \\
4.3 0 0 \\
};
\addlegendentry{\Large $P(H,\leq)$};

\addplot3[solid,fill=white!90!black,opacity=5.000000e-01,draw=black,forget plot]
table[row sep=crcr]{
x y z\\
0 0 0 \\
4.3 0 0 \\
4.3 0 0.5 \\
2.3 0 2.5 \\
0 0 2.5 \\
0 0 0 \\
};

\addplot3[solid,fill=white!90!black,opacity=4.000000e-01,draw=black,forget plot]
table[row sep=crcr]{
x y z\\
0 0 2.5 \\
2.3 0 2.5 \\
1 1.3 2.5 \\
0 1.3 2.5 \\
0 0 2.5 \\
};

\addplot3[solid,fill=white!90!black,opacity=4.000000e-01,draw=black,forget plot]
table[row sep=crcr]{
x y z\\
0 1.3 2.5 \\
1 1.3 2.5 \\
1 1.8 2 \\
0 1.8 2 \\
0 1.3 2.5 \\
};

\addplot3[solid,fill=white!90!black,opacity=4.000000e-01,draw=black,forget plot]
table[row sep=crcr]{
x y z\\
0 1.8 2 \\
1 1.8 2 \\
3 1.8 0 \\
0 1.8 0 \\
0 1.8 2 \\
};

\addplot3[solid,fill=white!90!black,opacity=4.000000e-01,draw=black,forget plot]
table[row sep=crcr]{
x y z\\
0 0 0 \\
0 0 2.5 \\
0 1.3 2.5 \\
0 1.8 2 \\
0 1.8 0 \\
0 0 0 \\
};

\addplot3[solid,fill=white!90!black,opacity=4.000000e-01,draw=black,forget plot]
table[row sep=crcr]{
x y z\\
0 1.8 0 \\
3 1.8 0 \\
4.3 0.5 0 \\
4.3 0 0 \\
0 0 0 \\
0 1.8 0 \\
};


\end{axis}
\end{tikzpicture}%
}
	\caption{The polyhedron $P(H,\leq)$ and base polyhedron $B(H,\leq)$ of the entropy function $H$ in the system in Example~\ref{ex:main}. Here, $B(H,\leq)$ coincides with the achievable rate region $\RRCO(V)$ for the lossless multiterminal data compression problem and the core $\Core{V}$ of the coalitional game $\G{V}{H}$. }
	\label{fig:Core}
\end{figure}
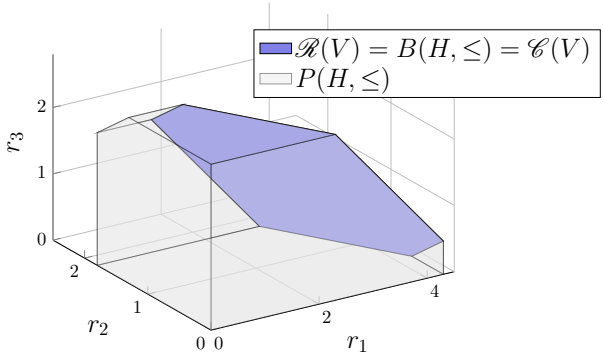

\begin{example} \label{ex:Core}
    Consider the system in Example~\ref{ex:main}. Fig.~\ref{fig:Core} shows $P(H,\leq)$ and $B(H,\leq)$, the polyhedron and base polyhedron of the entropy function $H$, respectively. The base polyhedron $B(H,\leq)$ coincides with the achievable rate region $\RRCO(V)$ and the core $\Core{V}$ of the game $\G{V}{H}$, which is a region on the plane $\Set{\rv_V \in \Real^{|V|} \colon r(V) = \frac{24}{5}}$ that is bounded by a set of linear constraints.
\end{example}

In general, the core $\Core{V}$ of a coalition game could be empty, i.e., there exists at least one user that is unwilling to cooperate in the grand coalition; If $\Core{V} \neq \emptyset$, it is not necessarily a singleton set. Then, the two fundamental questions in the coalitional game are:
\begin{enumerate}[(a)]
  \item Is the core nonempty, or do all the users in $V$ have incentive to cooperate in the grand coalition $V$?
  \item If the core is nonempty, can we find a fair cost/rate allocation method in the core $\Core{V}$?
\end{enumerate}
In the next subsection, we show that question (a) has a straightforward answer due to the convexity of the game $\G{V}{H}$. The Shapley value in Section~\ref{sec:Shapley} answers question (b).

\subsection{Convexity and Nonemptiness of Core}
\label{sec:Intepretation}

According to the definition in \cite[Section 2]{Shapley1971Convex}, a coalitional game is convex if and only if the characteristic cost function is submodular.\footnote{The definition in \cite[Section 2]{Shapley1971Convex} is based on the supermodularity of the characteristic payoff function, which corresponds to the submodularity of the characteristic cost function.}
Due to the fact that the entropy function $H$ is a polymatroid rank function, a subgroup of submodular function, it is straightforward that the game $\G{V}{H}$ is convex. The convexity of the game $\G{V}{H}$ is consistent with the results in \cite{Madiman2008ISIT,Madiman2008}.\footnote{The authors in \cite{Madiman2008ISIT,Madiman2008} formulated a coalitional game model $\G{V}{\Hdual}$ with the the dual set function $\Hdual$ being the characteristic payoff function, where the nonemptiness of the core $\Core{V}$ is due to the supermodularity of $\Hdual$. See Appendix~\ref{app:background} for the details. }
According to \cite[Theorem 4]{Shapley1971Convex}, the core of a convex game is nonempty. Therefore, for the multiterminal data compression problem, we have $\G{V}{H}$ being convex with a nonempty core $\Core{V}$, i.e., all the users would like to cooperate with each other and, therefore, the grand coalition $V$ forms necessarily.\footnote{On the other hand, the base polyhedron of a submodular function is nonempty \cite[Theorem 2.3]{Fujishige2005}. So, due to the equivalence \eqref{eq:Equivalence}, $\Core{V} \neq \emptyset$. Or, the nonemptiness of $\Core{V}$ is essentially due to the submodularity of $H$.}

\begin{example} \label{ex:CoopBehavior}
    For the system in Example~\ref{ex:main}, assume that the users are working individually at the beginning so that user $i$ encodes the source $\RZ{i}$ at the rate that is exactly equal to $H(\Set{i})$, i.e., $r_i = H(\Set{i})$ for all $i \in V$. However, it will not take long for user $i$ to realize that he/she can reduce the source coding rate by cooperating with another user. Take user $3$ for example: If user $3$ works alone, the source coding rate is $r_3 = H(\Set{3}) = \frac{5}{2}$; If user $3$ cooperates with user $2$, we have $r_2 + r_3 = r(\Set{2,3}) = H(\Set{2,3}) = \frac{19}{5}$, which means user $3$ can encode $\RZ{3}$ with a rate strictly lower than $\frac{5}{2}$ and so as user $2$, e.g., the rate $(r_2,r_3) = (\frac{3}{2},\frac{23}{10})$ is sufficient to encode $\RZ{\Set{2,3}}$. Note, the cooperative behavior of users $2$ and $3$ in coalition $\Set{2,3}$ is essentially due to the positive mutual information $I(\Set{2} \wedge \Set{3}) = H(\Set{2}) + H(\Set{3}) - H(\Set{2,3}) = \frac{5}{2} > 0$.

    After coalition $\Set{2,3}$ forms, users $2$ and $3$ will soon realize that, if they cooperate with user $1$, all the users can further reduce the source coding rates and therefore the grand coalition $V$ forms. In fact, by assuming that the users form any coalition $X \subsetneq V$, it can be shown that they would finally merge with others to form the grand coalition $V$.
\end{example}

Example~\ref{ex:CoopBehavior} reveals the reason for the nonemptiness of the core in the data compression problem: the nonnegativity of the mutual information. The mutual information can be considered as the information redundancy in the source coding, or data compression, problem, which can be minimized if the users cooperatively encode the multiple sources in $\RZ{V}$. The worst case is when the mutual information is zero, i.e., when all the sources in $\RZ{V}$ are mutually independent, where working separately or cooperating in $V$ are indifferent for the users. See the example below.

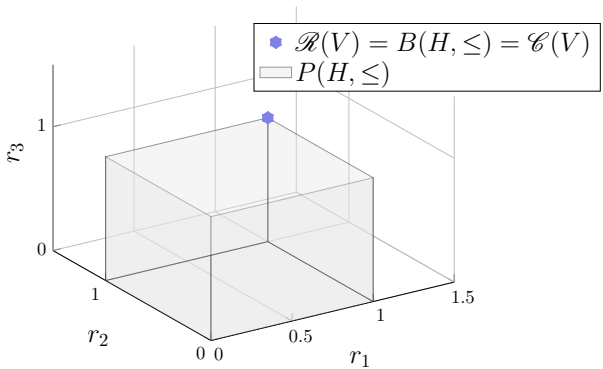
\begin{figure}[tbp]
	\centering
    \scalebox{0.7}{
%
%
%
\definecolor{mycolor1}{rgb}{1,1,0.5}%
\definecolor{mycolor2}{rgb}{0.5,0.5,0.9}%
\begin{tikzpicture}

\begin{axis}[%
width=3in,
height=2.5in,
view={-33}{30},
scale only axis,
xmin=0,
xmax=1.5,
xlabel={\Large $r_1$},
xmajorgrids,
ymin=0,
ymax=1.5,
ylabel={\Large $r_2$},
ymajorgrids,
zmin=0,
zmax=1.5,
zlabel={\Large $r_3$},
zmajorgrids,
axis x line*=bottom,
axis y line*=left,
axis z line*=left,
legend style={at={(0.5,0.95)},anchor=north west,draw=black,fill=white,legend cell align=left}
]

\addplot3 [
color=mycolor2,
line width=6.0pt,
only marks,
mark=asterisk,
mark options={solid}]
table[row sep=crcr] {
1 1 1\\
};
\addlegendentry{\Large $\RRCO(V) = B(H,\leq) = \Core{V}$};

\addplot3[area legend,solid,fill=white!90!black,opacity=4.000000e-01,draw=black]
table[row sep=crcr]{
x y z\\
0 0 0 \\
1 0 0 \\
1 1 0 \\
0 1 0 \\
0 0 0 \\
};

\addlegendentry{\Large $P(H,\leq)$};

\addplot3[solid,fill=white!90!black,opacity=4.000000e-01,draw=black,forget plot]
table[row sep=crcr]{
x y z\\
0 0 0 \\
0 1 0 \\
0 1 1 \\
0 0 1 \\
0 0 0 \\
};

\addplot3[solid,fill=white!90!black,opacity=4.000000e-01,draw=black,forget plot]
table[row sep=crcr]{
x y z\\
0 0 0 \\
1 0 0 \\
1 0 1 \\
0 0 1 \\
0 0 0 \\
};

\addplot3[solid,fill=white!90!black,opacity=4.000000e-01,draw=black,forget plot]
table[row sep=crcr]{
x y z\\
1 0 0 \\
1 1 0 \\
1 1 1 \\
1 0 1 \\
1 0 0 \\
};

\addplot3[solid,fill=white!90!black,opacity=4.000000e-01,draw=black,forget plot]
table[row sep=crcr]{
x y z\\
0 1 0 \\
0 1 1 \\
1 1 1 \\
1 1 0 \\
0 1 0 \\
};

\end{axis}
\end{tikzpicture}%
}
	\caption{The polyhedron $P(H,\leq)$ and base polyhedron $B(H,\leq)$ of the entropy function $H$ of a mutually independent multiple sources in Example~\ref{ex:IndependentSource}. In this case, $\Core{V} = \Set{(1,1,1)}$ and it makes no difference for the users to work separately or cooperate with each other. }
	\label{fig:CoreSingleton}
\end{figure}

\begin{example} \label{ex:IndependentSource}
    Assume the three users in $V = \Set{1,2,3}$ observe respectively
        \begin{equation}
            \begin{aligned}
                \RZ{1} &= (\RW{a}),  \\
                \RZ{2} &= (\RW{b}),  \\
                \RZ{3} &= (\RW{c}),
            \end{aligned} \nonumber
        \end{equation}
where $\RW{j}$ for all $j \in \Set{a,b,c}$ is an independent uniformly distributed random bit, i.e., the components in $\RZ{V}$ are mutually independent. Fig.~\ref{fig:CoreSingleton} shows the core $\Core{V}$ reduces to a point $\rv_V = (1,1,1)$. It can be shown that forming any coalition $X \subseteq V$, including the singleton and grand coalition, necessarily results in $r_i = 1$ for all $i \in V$, i.e., it makes no difference whether the users work separately or cooperate. Alternatively, encoding $\RZ{i}$s separately is equivalent to encoding $\RZ{V}$ cooperatively.
\end{example}

\section{Shapley Value}
\label{sec:Shapley}

Since the core is not a singleton in general, the question follows is how to select a cost allocation method, or source coding rate vector, $\rv_V$ in the core $\Core{V}$ of the game $\G{V}{H}$. In a coalitional game where the users are considered as peers, e.g., the sensor nodes in a WSN, a natural selection criterion is the fairness, i.e., we want to find a $\rv_V \in \Core{V}$ that distributes the total cost $H(V)$ as fairly as possible among the users in $V$. There is a well-known answer to this problem: The Shapley value $\rvS_V$ \cite{Shapley1953Value}, a unique value that lies in the core of the convex game $\G{V}{H}$, with each dimension being
\begin{equation} \label{eq:Shapley}
    \rS_i = \sum_{C \subseteq V \setminus \Set{i}} \frac{|C|! (|V| - |C| - 1)!}{|V|!} ( H(C \sqcup \Set{i}) - H(C)),
\end{equation}
where $\sqcup$ is the disjoint union. The Shapley value has been adopted in [7] as a fair source coding scheme. $\rvS$ is fair in that it assigns each player his/her expected marginal cost, which can be explained by its relationship with the extreme points in the core as follows.

\subsection{Extreme Points}

Let $\Phi_V = (\phi_1,\dotsc,\phi_{|V|})$ be a permutation of the user indices, e.g., $\Phi = (2,3,1,4)$ is a permutation of $V = \Set{1,\dotsc,4}$. Consider the Edmond greedy algorithm \cite{Edmonds2003Convex}: For $i = 1$ to $|V|$, do
    $$ r_{\phi_i} = H(\Set{\phi_i}|\Set{\phi_1,\dotsc,\phi_{i-1}})$$
with $r_{\phi_1} = H(\Set{\phi_1})$. The resulting $\rv_V$ is an extreme point in the core $\Core{V}$ \cite{Edmonds2003Convex}. In fact, the Edmond greedy algorithm states that an achievable rate vector in the core $\Core{V}$ can be reached if the users in $V$ encode the source with the side information at the sink in order. This approach has also been adopted in \cite{Schonberg2004} for the distributed source coding problem.
See also Example~\ref{ex:Edmond} below. Let $\EX(V)$ denote the set of all the extreme points in the core $\Core{V}$. $\EX(V)$ can be constructed by applying the Edmond greedy algorithm for all permutations of $V$.

\begin{figure}[tbp]
	\centering
    \scalebox{0.7}{
%
%
%
\definecolor{mycolor1}{rgb}{0.5,0.5,0.9}%
\begin{tikzpicture}

\begin{axis}[%
width=3in,
height=2.5in,
view={-33}{30},
scale only axis,
xmin=0,
xmax=4.5,
xlabel={\Large $r_1$},
xmajorgrids,
ymin=0,
ymax=2.5,
ylabel={\Large $r_2$},
ymajorgrids,
zmin=0,
zmax=2.8,
zlabel={\Large $r_3$},
zmajorgrids,
axis x line*=bottom,
axis y line*=left,
axis z line*=left,
legend style={at={(0.6,1.05)},anchor=north west,draw=black,fill=white,legend cell align=left}
]

\addplot3[area legend,solid,fill=mycolor1,draw=black]
table[row sep=crcr]{
x y z\\
4.3 0.5 0 \\
4.3 0 0.5 \\
2.3 0 2.5 \\
1 1.3 2.5 \\
1 1.8 2 \\
3 1.8 0 \\
};
\addlegendentry{\Large $\RRCO(V) = B(H,\leq) = \Core{V}$};

\addplot3[area legend,solid,fill=white!90!black,opacity=4.000000e-01,draw=black]
table[row sep=crcr]{
x y z\\
4.3 0 0 \\
4.3 0.5 0 \\
4.3 0 0.5 \\
4.3 0 0 \\
};
\addlegendentry{\Large $P(H,\leq)$};

\addplot3[solid,fill=white!90!black,opacity=5.000000e-01,draw=black,forget plot]
table[row sep=crcr]{
x y z\\
0 0 0 \\
4.3 0 0 \\
4.3 0 0.5 \\
2.3 0 2.5 \\
0 0 2.5 \\
0 0 0 \\
};

\addplot3[solid,fill=white!90!black,opacity=4.000000e-01,draw=black,forget plot]
table[row sep=crcr]{
x y z\\
0 0 2.5 \\
2.3 0 2.5 \\
1 1.3 2.5 \\
0 1.3 2.5 \\
0 0 2.5 \\
};

\addplot3[solid,fill=white!90!black,opacity=4.000000e-01,draw=black,forget plot]
table[row sep=crcr]{
x y z\\
0 1.3 2.5 \\
1 1.3 2.5 \\
1 1.8 2 \\
0 1.8 2 \\
0 1.3 2.5 \\
};

\addplot3[solid,fill=white!90!black,opacity=4.000000e-01,draw=black,forget plot]
table[row sep=crcr]{
x y z\\
0 1.8 2 \\
1 1.8 2 \\
3 1.8 0 \\
0 1.8 0 \\
0 1.8 2 \\
};

\addplot3[solid,fill=white!90!black,opacity=4.000000e-01,draw=black,forget plot]
table[row sep=crcr]{
x y z\\
0 0 0 \\
0 0 2.5 \\
0 1.3 2.5 \\
0 1.8 2 \\
0 1.8 0 \\
0 0 0 \\
};

\addplot3[solid,fill=white!90!black,opacity=4.000000e-01,draw=black,forget plot]
table[row sep=crcr]{
x y z\\
0 1.8 0 \\
3 1.8 0 \\
4.3 0.5 0 \\
4.3 0 0 \\
0 0 0 \\
0 1.8 0 \\
};

\addplot3 [
color=red,
line width=3.0pt,
only marks,
mark=triangle,
mark options={solid,,rotate=180}]
table[row sep=crcr] {
4.3 0.5 0\\
4.3 0 0.5\\
2.3 0 2.5\\
1 1.3 2.5\\
1 1.8 2\\
3 1.8 0\\
};
\addlegendentry{\Large $\EX(V)$};

\addplot3 [
->,
color=blue,
dashed,
mark=star,
mark options={solid,,rotate=180},
line width=2.0pt]
table[row sep=crcr] {
0 0 0\\
0 1.8 0\\
0 1.8 2\\
1 1.8 2\\
};
\addlegendentry{\Large path to $(1,\frac{9}{5},2)$};

\end{axis}
\end{tikzpicture}%
}
	\caption{The updated path of the source coding rate vector: $(0,0,0) \rightarrow (0,\frac{9}{5},0) \rightarrow (0,\frac{9}{5},2) \rightarrow (1,\frac{9}{5},2)$ resulting from the Edmond greedy algorithm \cite{Edmonds2003Convex} for the permutation $\Phi=(2,3,1)$ in the system in Example~\ref{ex:main}. The corresponding explanation is in Example~\ref{ex:Edmond}.}
	\label{fig:DemoEdmond}
\end{figure}
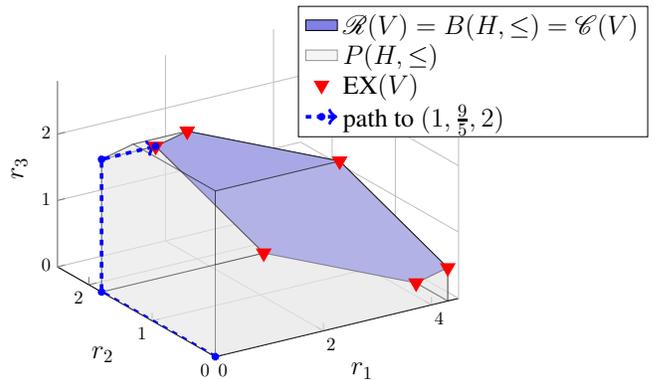

\begin{example} \label{ex:Edmond}
    We apply the Edmond greedy algorithm to the system in Example~\ref{ex:main} for the permutation $\Phi = (2,3,1)$. For $\phi_1 = 2$, we assign the source coding rate $r_2 = H(\Set{2}) = \frac{9}{5}$ so that the sink node $T$ obtain the source sequence $\RZ{2}^n$; For $\phi_2 = 3$, we assign the source coding rate to user $3$ conditioned on the information obtained by $T$, i.e., $r_3 = H(\Set{3}|\Set{T}) = H(\Set{3}|\Set{2}) = 2$ so that $H(\Set{T}) = H(\Set{2,3})$ and $\RZ{\Set{2,3}}^n$ is recovered at $T$; For $\phi_3 = 1$, user $3$ just need to send the remaining information, i.e., $r_1 = H(\Set{1}|\Set{T}) = H(\Set{1}|\Set{2,3}) = 1$ so that $\RZ{\Set{1,2,3}}^n$ is recovered at $T$. The resulting rate vector $\rv_V = (1,\frac{9}{5},2)$ is an extreme point in the core $\Core{V}$. See Fig.~\ref{fig:DemoEdmond}. We have the extreme point set \begin{equation} \label{eq:EX}
        \begin{aligned}
            \EX(V) = \Big\{ & (\frac{43}{10},\frac{1}{2},0), (\frac{43}{10},0,\frac{1}{2}), (3,\frac{9}{5},0),\\
                                    & (1,\frac{9}{5},2), (\frac{23}{10},0,\frac{5}{2}), (1,\frac{13}{10},\frac{5}{2}) \Big\},
        \end{aligned}
    \end{equation}
    where all $\rv_V \in \EX(V)$ can be determined by applying the Edmond greedy algorithm for all $|V|! = 6$ permutations of $V$.
\end{example}

The Shapley value assigns each user $i$ the expected source coding rate $\rS_i$ resulting from the Edmond greedy algorithm by assuming that each order, or permutation, $\Phi$ to form the grand coalition is equiprobable.
For a subset $C \subsetneq V$ such that $i \notin C$, assume that the user $i$ joins the coalition $C \sqcup \Set{i}$ after the coalition $C$ is formed and, then, the rest of the users in $V \setminus (C \sqcup \Set{i}) $ join and form the grand coalition $V$. In the Edmond greedy algorithm, user $i$ will be assigned the rate $H(C \sqcup \Set{i}) - H(C)$ for the $ |C|! (|V| - |C| - 1)!$ out of $|V|!$ time, which explains the weight in \eqref{eq:Shapley}. In other words, the Shapley value $\rvS_V$ is the average value over all the extreme points in the core $\Core{V}$:
$$ \rvS_V = \frac{\sum_{\rv_V\EX(V)} \rv_V}{|\EX(V)|} $$
This is the reason that $\rvS_V$ is called the gravity center of the extreme points of the core in a convex game in \cite{Shapley1953Value}.

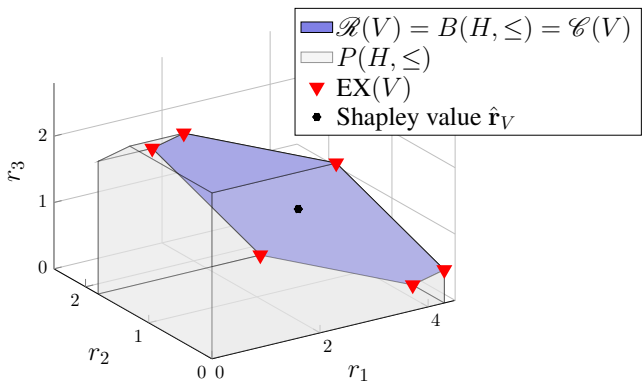
\begin{figure}[tbp]
	\centering
    \scalebox{0.7}{
%
%
%
\definecolor{mycolor1}{rgb}{0.5,0.5,0.9}%
\begin{tikzpicture}

\begin{axis}[%
width=3in,
height=2.5in,
view={-33}{30},
scale only axis,
xmin=0,
xmax=4.5,
xlabel={\Large $r_1$},
xmajorgrids,
ymin=0,
ymax=2.5,
ylabel={\Large $r_2$},
ymajorgrids,
zmin=0,
zmax=2.8,
zlabel={\Large $r_3$},
zmajorgrids,
axis x line*=bottom,
axis y line*=left,
axis z line*=left,
legend style={at={(0.6,1.05)},anchor=north west,draw=black,fill=white,legend cell align=left}
]

\addplot3[area legend,solid,fill=mycolor1,draw=black]
table[row sep=crcr]{
x y z\\
4.3 0.5 0 \\
4.3 0 0.5 \\
2.3 0 2.5 \\
1 1.3 2.5 \\
1 1.8 2 \\
3 1.8 0 \\
};
\addlegendentry{\Large $\RRCO(V) = B(H,\leq) = \Core{V}$};

\addplot3[area legend,solid,fill=white!90!black,opacity=4.000000e-01,draw=black]
table[row sep=crcr]{
x y z\\
4.3 0 0 \\
4.3 0.5 0 \\
4.3 0 0.5 \\
4.3 0 0 \\
};
\addlegendentry{\Large $P(H,\leq)$};

\addplot3[solid,fill=white!90!black,opacity=5.000000e-01,draw=black,forget plot]
table[row sep=crcr]{
x y z\\
0 0 0 \\
4.3 0 0 \\
4.3 0 0.5 \\
2.3 0 2.5 \\
0 0 2.5 \\
0 0 0 \\
};

\addplot3[solid,fill=white!90!black,opacity=4.000000e-01,draw=black,forget plot]
table[row sep=crcr]{
x y z\\
0 0 2.5 \\
2.3 0 2.5 \\
1 1.3 2.5 \\
0 1.3 2.5 \\
0 0 2.5 \\
};

\addplot3[solid,fill=white!90!black,opacity=4.000000e-01,draw=black,forget plot]
table[row sep=crcr]{
x y z\\
0 1.3 2.5 \\
1 1.3 2.5 \\
1 1.8 2 \\
0 1.8 2 \\
0 1.3 2.5 \\
};

\addplot3[solid,fill=white!90!black,opacity=4.000000e-01,draw=black,forget plot]
table[row sep=crcr]{
x y z\\
0 1.8 2 \\
1 1.8 2 \\
3 1.8 0 \\
0 1.8 0 \\
0 1.8 2 \\
};

\addplot3[solid,fill=white!90!black,opacity=4.000000e-01,draw=black,forget plot]
table[row sep=crcr]{
x y z\\
0 0 0 \\
0 0 2.5 \\
0 1.3 2.5 \\
0 1.8 2 \\
0 1.8 0 \\
0 0 0 \\
};

\addplot3[solid,fill=white!90!black,opacity=4.000000e-01,draw=black,forget plot]
table[row sep=crcr]{
x y z\\
0 1.8 0 \\
3 1.8 0 \\
4.3 0.5 0 \\
4.3 0 0 \\
0 0 0 \\
0 1.8 0 \\
};

\addplot3 [
color=red,
line width=3.0pt,
only marks,
mark=triangle,
mark options={solid,,rotate=180}]
table[row sep=crcr] {
4.3 0.5 0\\
4.3 0 0.5\\
2.3 0 2.5\\
1 1.3 2.5\\
1 1.8 2\\
3 1.8 0\\
};
\addlegendentry{\Large $\EX(V)$};

\addplot3 [
color=black,
line width=3.0pt,
only marks,
mark=asterisk,
mark options={solid}]
table[row sep=crcr] {
2.65 0.9 1.25\\
};
\addlegendentry{\Large Shapley value $\rvS_V$};

%

%

\end{axis}
\end{tikzpicture}%
}
	\caption{The Shapley value $\rvS_V = (\frac{53}{20},\frac{9}{10},\frac{5}{4})$ for the system in Example~\ref{ex:main}. $\rvS_V$ is a fair rate allocation method in the core $\Core{V}$. In fact, $\rvS_V$ is the gravity center of all the extreme points in $\EX(V)$. }
	\label{fig:Shapley}
\end{figure}

\begin{example} \label{ex:Shapley}
    For the system in Example~\ref{ex:main}, we have the Shapley value in core $\Core{V}$ being $\rvS_V = (\frac{53}{20},\frac{9}{10},\frac{5}{4})$ in Fig.~\ref{fig:Shapley}, which is exactly the value of $\frac{\sum_{\rv_V\EX(V)}\rv_V}{|\EX(V)|}$, where the extreme point set $\EX(V)$ is enumerated in \eqref{eq:EX}.
\end{example}

\subsection{Decomposition}
\label{sec:Decomposition}

Although the Shapley value is the desired fair cost allocation method in the core $\Core{V}$, calculating it might be cumbersome in large scale systems. Assume that the value of $H(X)$ for any $X \subseteq V$ can be obtained by an oracle call and $\delta$ refers to the upper bound on the computation time of this oracle call. Obtaining the Shapley value in \eqref{eq:Shapley} requires the value of $H(X)$ for all $X \subseteq V$ so that the complexity is $O(2^{|V|} \cdot \delta)$.\footnote{We assume that the time complexity of the multiplication and summation operations in \eqref{eq:Shapley} is much less than that of the oracle call of $H$.}
Then, there is a problem of how to alleviate this exponentially growing complexity in large $|V|$.
On the other hand, when the system size $|V|$ is large, it is very possible that the coalitional game is decomposable: The users' cooperating in smaller coalitions is equivalent to cooperating in the grand coalition. In this section, we show that the calculation of the Shapley value in a decomposable game is separable, which results in a reduction in computational complexity.

\begin{definition}[Decomposable Convex Game {\cite[Theorems 3.32 and 3.38, Lemma 3.37]{Fujishige2005}}\footnote{This definition is based on the concept of the separator of a disconnected submodular system in \cite{Fujishige2005,Bilxby1985}, where the minimal separators are also called the elementary separators in \cite[Section 3]{Bilxby1985} and the principal partition of a polymatroid in \cite{FujishigePincStruct}.}] \label{def:DecomposeGame}
    The convex game $\G{V}{H}$ is decomposable if
    $$ H(V) = \sum_{C \in \Pat} H(C),$$
    for some proper partition $\Pat$ of user set $V$. Here, $\Pat$ is called a decomposer and each $C \in \Pat$ forms a subgame $\G{C}{H}$ that is convex. A game $\G{V}{H}$ is indecomposable if it has only one decomposer $\Set{V}$.
\end{definition}

It is easy to see that, for any decomposer $\Pat$ of a decomposable game $\G{V}{H}$, the components in $(\RZ{C} \colon C \in \Pat)$ are mutually independent. For any $X, Y \subset V$ such that $X \cap Y = \emptyset$, let $\rv_X \oplus \rv_Y = \rv_{X \sqcup Y}$ be the \textit{direct sum} of $\rv_X$ and $\rv_Y$. For example, for $\rv_{\Set{1,3}} = (r_1,r_3) = (3,7)$ and $\rv_{\Set{2,5,6}} = (r_2,r_5,r_6) = (5,2,4)$, $\rv_{\Set{1,3}} \oplus \rv_{\Set{2,5,6}} = \rv_{\Set{1,2,3,5,6}} = (3,5,7,2,4)$. We have a decomposable core $\Core{V}$ for a decomposable game $\G{V}{H}$.

\begin{lemma}[{\cite[Theorems 3.15, 3.32 and 3.38, Corollary 3.40]{Fujishige2005}, \cite[Theorem 6]{Shapley1971Convex}}]  \label{lemma:MinSepProp}
    For a decomposer $\Pat$ of a decomposable convex game $\G{V}{H}$, the core $\Core{V}$ is decomposable:
                \begin{equation}
                    \begin{aligned}
                        \Core{V} & = \oplus_{C \in \Pat} \Core{C} \\
                                         & = \Set{ \oplus_{C \in \Pat} \rv_C \colon  \rv_C \in \Core{C}, C \in \Pat },
                    \end{aligned} \nonumber
                \end{equation}
    where $\Core{C}$ is the core of the subgame $\G{C}{H}$; $\Core{V}$ has dimension $|V| - |\Pat^*|$ where $\Pat^*$ is the finest decomposer.\footnote{The decomposition of $\Core{V}$ was originally derived in \cite[Theorems 3.15 and 3.38, Corollary 3.40]{Fujishige2005} as the separation of the base polyhedron in a disconnected submodular system. We express it in terms of $\Core{V}$ in Lemma~\ref{lemma:MinSepProp} due to the equivalence~\eqref{eq:Equivalence}. Also, for a decomposable convex game, the finest decomposer uniquely exists. See Appendix~\ref{app:background}. For an indecomposable game $\G{V}{H}$, the core $\Core{V}$ has the full dimension $|V| - 1$ \cite[Theorem 6(a)]{Shapley1971Convex} since $\Pat^* =\Set{V}$ is the only one decomposer.} \hfill \IEEEQED
\end{lemma}

We also have the decomposable Shapley value $\rvS_V$ due to the decomposition of the core $\Core{V}$ as follows. The proof of Theorem~\ref{theo:DecompShapley} is in Appendix~\ref{app:mainproof}.

\begin{theorem} \label{theo:DecompShapley}
    If $\G{V}{H}$ is decomposable, we have
    $$ \rvS_V = \oplus_{C \in \Pat} \rvS_C $$
    for all decomposers $\Pat$, where $\rvS_C$ is the Shapley value in the core $\Core{C}$ of subgame $\G{C}{H}$. \hfill\IEEEQED
\end{theorem}

\begin{remark}
    Theorem~\ref{theo:DecompShapley} is a recast of \cite[Corollary 2]{Shapley1953Value}, which is derived in terms of the characteristic payoff function based on the concept of carriers and is proved by the the additivity of the Shapley value in \cite[Axiom 3]{Shapley1953Value}.\footnote{Based on the definition in \cite[Section 2]{Shapley1953Value}, a carrier corresponds to a coalition $C$ in a decomposer $\Pat$. The additivity in \cite[Axiom 3]{Shapley1953Value} is not a property specifically for the decomposable game. Therefore, the application of \cite[Corollary 2]{Shapley1953Value} to decomposable games is not explicit in \cite{Shapley1953Value}. }
    But, Theorem~\ref{theo:DecompShapley} is derived for the decomposable game in terms of the characteristic cost function and the proof is based on the decomposition of the core $\Core{V}$ in Lemma~\ref{lemma:MinSepProp}.
\end{remark}

According to Theorem~\ref{theo:DecompShapley}, the Shapley value $\rvS_V$ in a decomposable game $\G{V}{H}$ can be obtained separately for each subgame $\G{C}{H}$ for all coalitions $C$ in a decomposer $\Pat$. In Section~\ref{sec:Complexity}, we show the advantage of this separation: the reduction in computational complexity.

\begin{example}\label{ex:Decompose}
    For the system in Example~\ref{ex:main}, the game $\G{V}{H}$ is indecomposable since $\Set{V}$ is the only decomposer. Therefore, we have $|V| - |\Set{V}| = 2$ and the core $\Core{V}$ in Fig.~\ref{fig:Core} is a $2$-dimensional plane. For the system in Example~\ref{ex:IndependentSource}, the game $\G{V}{H}$ is decomposable with all partitions $\Pat$ of $V$ being the decomposer and the finest one is $\Pat^* = \Set{\Set{1},\Set{2},\Set{3}}$. Therefore, the dimension of the core $\Core{V}$ in Fig.~\ref{fig:CoreSingleton} is $|V| - |\Pat^*| = 0$: $\Core{V}$ reduces to a point $(1,1,1)$ which is necessarily the Shapley value.
\end{example}

\begin{figure}[tbp]
	\centering
    \scalebox{0.7}{
%
%
%
\definecolor{mycolor1}{rgb}{0.5,0.5,0.9}%
\begin{tikzpicture}

\begin{axis}[%
width=3in,
height=2.5in,
view={-35}{30},
scale only axis,
xmin=0,
xmax=2.3,
xlabel={\Large $r_1$},
xmajorgrids,
ymin=0,
ymax=1.3,
ylabel={\Large $r_2$},
ymajorgrids,
zmin=0,
zmax=1.9,
zlabel={\Large $r_3$},
zmajorgrids,
axis x line*=bottom,
axis y line*=left,
axis z line*=left,
legend style={at={(0.6,1.05)},anchor=north west,draw=black,fill=white,legend cell align=left}
]

\addplot3[
area legend,
color=mycolor1,
line width=5.0pt]
table[row sep=crcr]{
x y z\\
1 1 1.6 \\
2 1 0.6 \\
};
\addlegendentry{\Large $\RRCO(V) = B(H,\leq) = \Core{V}{H}$};


\addplot3[area legend,solid,fill=white!90!black,opacity=4.000000e-01,draw=black]
table[row sep=crcr]{
x y z\\
0 0 1.6 \\
1 0 1.6 \\
1 1 1.6 \\
0 1 1.6 \\
0 0 1.6 \\
};
\addlegendentry{\Large $P(H,\leq)$};

\addplot3[solid,fill=white!90!black,opacity=4.000000e-01,draw=black,forget plot]
table[row sep=crcr]{
x y z\\
0 0 0 \\
0 0 1.6 \\
1 0 1.6 \\
2 0 0.6 \\
2 0 0 \\
0 0 0 \\
};

\addplot3[solid,fill=white!90!black,opacity=4.000000e-01,draw=black,forget plot]
table[row sep=crcr]{
x y z\\
1 0 1.6 \\
1 1 1.6 \\
2 1 0.6 \\
2 0 0.6 \\
1 0 1.6 \\
};

\addplot3[solid,fill=white!90!black,opacity=4.000000e-01,draw=black,forget plot]
table[row sep=crcr]{
x y z\\
2 0 0 \\
2 0 0.6 \\
2 1 0.6 \\
2 1 0 \\
2 0 0 \\
};

\addplot3[solid,fill=white!90!black,opacity=4.000000e-01,draw=black,forget plot]
table[row sep=crcr]{
x y z\\
0 1 0 \\
2 1 0 \\
2 1 0.6 \\
1 1 1.6 \\
0 1 1.6 \\
0 1 0 \\
};

\addplot3[solid,fill=white!90!black,opacity=4.000000e-01,draw=black,forget plot]
table[row sep=crcr]{
x y z\\
0 0 0 \\
0 0 1.6 \\
0 1 1.6 \\
0 1 0 \\
0 0 0 \\
};

\addplot3[solid,fill=white!90!black,opacity=4.000000e-01,draw=black,forget plot]
table[row sep=crcr]{
x y z\\
0 0 0 \\
2 0 0 \\
2 1 0 \\
0 1 0 \\
0 0 0 \\
};

\addplot3 [
color=red,
line width=3.0pt,
only marks,
mark=triangle,
mark options={solid,,rotate=180}]
table[row sep=crcr] {
1 1 1.6\\
2 1 0.6\\
};
\addlegendentry{\Large $\EX(V)$};

\addplot3 [
color=black,
line width=3.0pt,
only marks,
mark=asterisk,
mark options={solid}]
table[row sep=crcr] {
1.5 1 1.1\\
};
\addlegendentry{\Large Shapley value $\rvS_V$};

\end{axis}
\end{tikzpicture}%
}
	\caption{The core $\Core{V}$ of the decomposable game $\G{V}{H}$ in the system in Example~\ref{ex:DecomposeSpecial}. The finest decomposer is $\Pat^* = \Set{\Set{1,3},\Set{2}}$ so that the dimension of $\Core{V}$ is $|V| - |\Pat^*| = 1$ and the Shapley value $\rvS_V = \rvS_{\Set{1,3}} \oplus \rS_2 = (\frac{3}{2},1,\frac{11}{10})$.}
	\label{fig:DecomposeSpeical}
\end{figure}
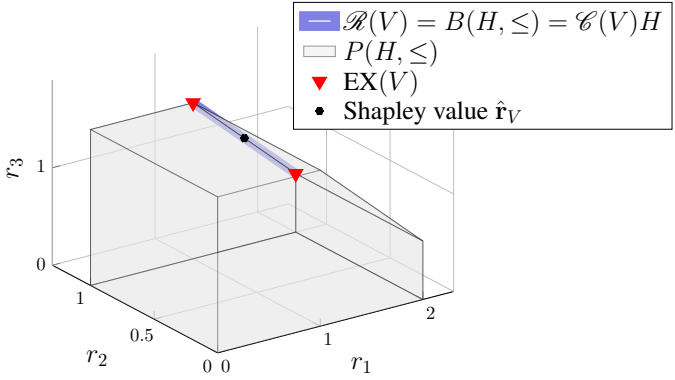

\begin{example}\label{ex:DecomposeSpecial}
    Assume that the users in $V = \Set{1,2,3}$ observe respectively
        \begin{equation}
            \begin{aligned}
                \RZ{1} &= (\RW{a},\RW{b}),  \\
                \RZ{2} &= (\RW{d}),  \\
                \RZ{3} &= (\RW{b},\RW{c}),
            \end{aligned} \nonumber
        \end{equation}
    where $\RW{c}$ is an independent random bit with $H(\RW{c}) = \frac{3}{5}$ and all other $\RW{j}$ are independent uniformly distributed random bit. In this system, the game is decomposable with the finest decomposer $\Pat^* = \Set{\Set{1,3},\Set{2}}$. It is easy to see why $\Pat^* = \Set{\Set{1,3},\Set{2}}$: $\RZ{\Set{1,3}}$ and $\RZ{2}$ are mutually independent. The core $\Core{V}$ in Fig.~\ref{fig:DecomposeSpeical} has the dimension $|V| - |\Pat^*| = 1$ and reduces to a line segment.
    We calculate the Shapley value separately for each subgame $\G{C}{H}$: $\rvS_{\Set{1,3}} = (\rS_1,\rS_3) = (\frac{3}{2},\frac{11}{10})$ for $\Set{1,3}$ and $\rS_2 = 1 $ for $\Set{2}$. Then, $\rvS_{\Set{1,3}} \oplus \rS_2 = (\rS_1,\rS_2,\rS_3) = (\frac{3}{2},1,\frac{11}{10})$ is the Shapley value of game $\G{V}{H}$.
\end{example}

\subsection{Determining Finest Decomposer}

To utilize the decomposition property of a decomposer, the first question is how to determine a decomposer. In real applications, the decomposer would not be as obvious as in Example~\ref{ex:DecomposeSpecial}. In fact, we would not have any prior knowledge of the multiple random sources. 
In this case, whether or not the game $\G{V}{H}$ is decomposable and the finest decomposer $\Pat^*$, if $\G{V}{H}$ is decomposable, can be determined by Algorithm~\ref{algo:Decomposer} that completes in $O(|V|^2 \cdot \delta)$ time, which does not require the knowledge of $\RZ{V}$.
Note, Algorithm~\ref{algo:Decomposer} also returns an extreme point in the core $\rv_V \in \Core{V}$ since steps 2, 4 and 5 are in fact implementing the Edmond greedy algorithm.

\begin{example}\label{ex:Decomposer}
    For the system in Example~\ref{ex:DecomposeSpecial}, by applying Algorithm~\ref{algo:Decomposer} for the permutation $\Phi = (3,2,1)$, we have $\rv_V = (1,1,\frac{8}{5})$, which is an extreme point in the core $\Core{V}$ as shown in Fig.~\ref{fig:DecomposeSpeical}. We also have $\hat{X}_1 = \Set{1,3}$, $\hat{X}_2 = \Set{2}$ and $\hat{X}_3 = \Set{3}$ at the end of each iteration. Since $\hat{X}_1$ and $\hat{X}_3$ are intersecting, they are merged and result in the finest decomposer $\Pat^* = \Set{\Set{1,3},\Set{2}}$.

    Consider the indecomposable game in Example~\ref{ex:main}. For the permutation $\Phi = (2,3,1)$, Algorithm~\ref{algo:Decomposer} returns an extreme point $\rv_V = (1,\frac{9}{5},2)$. We also have $\hat{X}_1 = \Set{1,2,3}$, $\hat{X}_2 = \Set{2}$ and $\hat{X}_3 = \Set{2,3}$, all of which are merged so that $\Pat^* = \Set{\Set{1,2,3}}$, which means the game is indecomposable.
\end{example}

\subsection{Complexity}
\label{sec:Complexity}

For the finest decomposer $\Pat^*$ of a decomposable convex game $\G{V}{H}$, let $\hat{C} = \arg\max \Set{ |C| \colon C \in \Pat^*}$ be the maximum size of the subgame. The separable calculation of the Shapley value $\oplus_{C \in \Pat^*}\rvS_C$ according to Theorem~\ref{theo:DecompShapley} reduces the complexity from $O(2^{|V|} \cdot \delta)$ to $O(\frac{|V|}{|\hat{C}|} \cdot 2^{|\hat{C}|} \cdot \delta)$. We remark that $O(\frac{|V|}{|\hat{C}|} \cdot 2^{|\hat{C}|} \cdot \delta)$ is an upper bound in that (a) obtaining the Shapley value $\rvS_{\hat{C}}$ for subgame $\G{\hat{C}}{H}$ completes in $O(2^{|\hat{C}|} \cdot \delta)$ time and (b) there are at most $\frac{|V|}{|\hat{C}|}$ such subgames.
For example, for the system in Example~\ref{ex:DecomposeSpecial}, if obtaining the Shapley value for each element in the finest decomposer $\Pat^*=\Set{\Set{1,3},\Set{2}}$, we can reduce the $2^3=8$ oracle calls of $H$ to $2^2+2 = 6$.

        \begin{algorithm} [t]
	       \label{algo:Decomposer}
	       \small
	       \SetAlgoLined
	       \SetKwInOut{Input}{input}\SetKwInOut{Output}{output}
	       \SetKwFor{For}{for}{do}{endfor}
            \SetKwRepeat{Repeat}{repeat}{until}
            \SetKwIF{If}{ElseIf}{Else}{if}{then}{else if}{else}{endif}
	       \BlankLine
           \Input{a convex coalitional game $\G{V}{H}$}
	       \Output{$\Pat^*$, the finest decomposer, if $\G{V}{H}$ is decomposable or $\Pat^* = \Set{V}$ if $\G{V}{H}$ is indecomposable and an extreme point $\rv_V \in \Core{V}$}
	       \BlankLine
            choose any permutation $\Phi$\;
            $r_{\phi_1} \leftarrow H(\Set{\phi_1})$ and $\hat{X}_{\phi_1} \leftarrow \Set{\phi_1}$\;

            \For{$i=2$ \emph{\KwTo} $|V|$}{
                $\hat{X}_{\phi_i} \leftarrow \Set{\phi_1, \dotsc, \phi_i}$\;
                $r_{\phi_i} \leftarrow H(\hat{X}_{\phi_i}) - H(\hat{X}_{\phi_i} \setminus \Set{\phi_i})$\;
                \For{$j=1$ \emph{\KwTo} $i-1$}{
                \If{$r(\hat{X}_{\phi_i} \setminus \Set{\phi_{i-j}}) = H(\hat{X}_{\phi_i} \setminus \Set{\phi_{i-j}}) $}{
                    $\hat{X}_{\phi_i} \leftarrow \hat{X}_{\phi_i} \setminus \Set{\phi_{i-j}}$ \; }
                }
            }
            $\Pat^* \leftarrow \Set{\hat{X}_{\phi_i} \colon i \in V}$ and keep merging any two intersecting elements in $\Pat^*$ until there is no left\;
            return $\Pat^*$ and $\rv_V$\;
	   \caption[Finest Decomposer ]{Finest Decomposer \cite[Algorithm 2.6]{Bilxby1985}\footnotemark}
	   \end{algorithm}
        \footnotetext{Algorithm~\ref{algo:Decomposer} is adapted from \cite[Algorithm 2.6]{Bilxby1985}. It has also been independently proposed in \cite{FujishigePincStruct} \cite[Secton 3.3]{Fujishige2005}. The finest decomposer has different names: It is called the minimal separators of a polymatroid in \cite[Section 3]{Bilxby1985}, the principal partition of a polymatroid in \cite[Section 3(b)]{FujishigePincStruct} and the finest partition of a disconnected submodular system in \cite[Secton 3.3]{Fujishige2005}. The validity of Algorithm~\ref{algo:Decomposer} is due to the lattice structure of the minimizer set of the problem $\min \Set{H(X) - r(X) \colon i \in X \subseteq V}$ for all $i \in V$ and $\rv_V \in B(H,\leq)$. Please see \cite{FujishigePincStruct} for further details.}

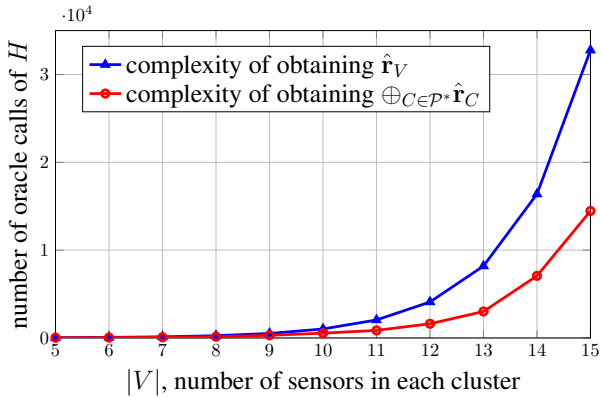
\begin{figure}[tbp]
	\centering
    \scalebox{0.7}{
%
%
\begin{tikzpicture}

\begin{axis}[%
width=4in,
height=2.3in,
scale only axis,
xmin=5,
xmax=15,
xlabel={\Large $|V|$, number of sensors in each cluster},
xmajorgrids,
ymin=0,
ymax=35000,
ylabel={\Large number of oracle calls of $H$},
ymajorgrids,
legend style={at={(0.05,0.95)},anchor=north west,draw=black,fill=white,legend cell align=left}
]

\addplot [
color=blue,
solid,
line width = 1.5,
mark=triangle,
mark options={solid},
]
table[row sep=crcr]{
5 32\\
6 64\\
7 128\\
8 256\\
9 512\\
10 1024\\
11 2048\\
12 4096\\
13 8192\\
14 16384\\
15 32768\\
};
\addlegendentry{\Large complexity of obtaining $\rvS_V$};

\addplot [
color=red,
solid,
line width = 1.5,
mark=o,
mark options={solid},
]
table[row sep=crcr]{
5 27.4\\
6 41.8\\
7 75.4\\
8 139.2\\
9 303\\
10 543.6\\
11 862.6\\
12 1617\\
13 3012.8\\
14 7070.8\\
15 14458.6\\
};
\addlegendentry{\Large complexity of obtaining $\oplus_{C \in \Pat^*} \rvS_C$};

\end{axis}
\end{tikzpicture}%
}
	\caption{The complexity comparison between the two methods for obtaining the Shapley value in decomposable games in Experiment~\ref{exp:ShapleyCompare}: obtaining $\rvS_V$ in \eqref{eq:Shapley} and $\oplus_{C \in \Pat^*}\rvS_C$ for the finest decomposer $\Pat^*$ according to Theorem~\ref{theo:DecompShapley}. Note, the complexity of obtaining $\oplus_{C \in \Pat^*}\rvS_C$ includes the number of oracle calls of $H$ for determining $\Pat^*$ in Algorithm~\ref{algo:Decomposer}.}
	\label{fig:Complex}
\end{figure}

\begin{experiment} \label{exp:ShapleyCompare}
    Consider the WSN in Fig.~\ref{fig:WSN}. We set the number of clusters to $20$. For each cluster, we fix $H(V) = 50$ and vary the number of sensors $|V|$ from $5$ to $15$. For each value of $|V|$, we do the followings for each cluster:
    \begin{enumerate}[(a)]
      \item randomly generate random sources $\RZ{V}$ so that the game $\G{V}{H}$ is decomposable;
      \item obtain the Shapley value $\rvS_V$ in \eqref{eq:Shapley} and record the number of calls of $H$;
      \item run Algorithm~\ref{algo:Decomposer} to determine the finest decomposer $\Pat^*$ and obtain the Shapley value $\rvS_C$ for each $C \in \Pat^*$ according to Theorem~\ref{theo:DecompShapley}. Record the number of calls of $H$ in both Algorithm~\ref{algo:Decomposer} and obtaining $\oplus_{C \in \Pat^*}\rvS_C$.
    \end{enumerate}
    We plot the mean number of oracle calls of $H$ of two methods over clusters in Fig.~\ref{fig:Complex}. It can be seen that the separate calculation of the Shapley value in a decomposable convex game $\G{V}{H}$ contributes to a considerable reduction in computational complexity as $|V|$ grows.
\end{experiment}

It should be noted that overall $O(\frac{|V|}{|\hat{C}|} \cdot 2^{|\hat{C}|} \cdot \delta)$ oracle calls can be distributed among the subgames $\G{C}{H}$. And, the computations of $\rvS_C$ can be done in a decentralized manner.
For example, for the system in Example~\ref{ex:DecomposeSpecial}, the users in $\Set{1,3}$ and user $2$ can calculate $\rvS_{\Set{1,3}}$ and $\rS_2$, respectively, in parallel.

\begin{figure}[tbp]
	\centering
    \scalebox{0.7}{
%
%
\begin{tikzpicture}

\begin{axis}[%
width=4in,
height=2.3in,
scale only axis,
xmin=5,
xmax=15,
xlabel={\Large $|V|$, number of sensors in each cluster},
xmajorgrids,
ymin=0,
ymax=1.2,
ylabel={\Large completion time in seconds},
ymajorgrids,
legend style={at={(0.025,0.95)},anchor=north west,draw=black,fill=white,legend cell align=left}
]

\addplot [
color=blue,
solid,
line width = 1.5,
mark=triangle,
mark options={solid},
]
table[row sep=crcr]{
5 0.00159142642716015\\
6 0.00104697187726324\\
7 0.00239370372774501\\
8 0.00583474400425353\\
9 0.0110973184245899\\
10 0.0241828258686293\\
11 0.0735972728406524\\
12 0.124015295781407\\
13 0.256420588769431\\
14 0.575262727582367\\
15 1.19120529036239\\
};
\addlegendentry{\Large completion time of calculating $\rvS_V$};

\addplot [
color=red,
solid,
line width = 1.5,
mark=o,
mark options={solid},
]
table[row sep=crcr]{
5 0.000223377869890115\\
6 0.000270984076652093\\
7 0.000679780828450106\\
8 0.00142811989895019\\
9 0.00404463791335763\\
10 0.00770753106984606\\
11 0.0261203255495417\\
12 0.0482743844063284\\
13 0.0999003651157365\\
14 0.176614418965252\\
15 0.419761396100415\\
};
\addlegendentry{\Large completion time of calculating $\oplus_{C \in \Pat^*} \rvS_C$};

\end{axis}
\end{tikzpicture}%
}
	\caption{The completion time comparison between the two methods for obtaining the Shapley value in decomposable games in Experiment~\ref{exp:ShapleyParallel}: calculating $\rvS_V$ in \eqref{eq:Shapley} and $\oplus_{C \in \Pat^*}\rvS_C$ for the finest decomposer $\Pat^*$ according to Theorem~\ref{theo:DecompShapley}. Here, $\rvS_C$ is calculated by each subgame $\G{C}{H}$ in parallel. We assume that all values of $H$ have been obtained in Experiment~\ref{exp:ShapleyParallel}, i.e., the completion time does not include the oracle calls of $H$.}
	\label{fig:CompletionTime}
\end{figure}
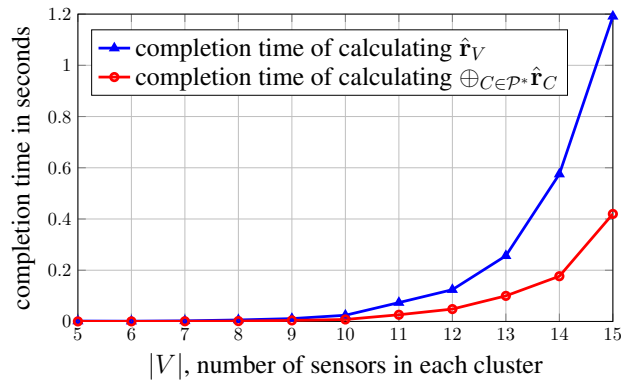

\begin{experiment} \label{exp:ShapleyParallel}
    In Experiment~\ref{exp:ShapleyCompare}, we allow the subgames $\G{C}{H}$ for all $C \in \Pat^*$ in each cluster to obtain $\rvS_C$ in parallel so that the completion time of calculating $\oplus_{C \in \Pat^*}\rvS_C$ is determined by $|\hat{C}|$, the maximum size of subgames. After obtaining the values of $H$ by the oracle calls in Experiment~\ref{exp:ShapleyCompare}, the average run time for calculating the values of $\rvS_V$ and $\oplus_{C \in \Pat^*}\rvS_C$ over clusters is shown in Fig.~\ref{fig:CompletionTime}.\footnote{We run Experiment~\ref{exp:ShapleyParallel} in MATLAB 2017a by a desktop computer with Intel Core i7-6600U processer, 8Gb RAM and 64-bit Windows 10 Enterprise operating system.}
    It can be shown that it is much faster to get $\oplus_{C \in \Pat^*}\rvS_C$ by allowing parallel computation.
\end{experiment}

In addition to the decomposition property, some Shapley value approximation methods, e.g., \cite{LibenNowell2012,Conitzer2004,Fatima2008}, can be applied to each subgame $\G{C}{H}$ to further reduce the complexity. For example, based on the random permutation method in \cite{LibenNowell2012}, random samples of the permutations of $C$, with the size quadratically growing in $|C|$, are generated so that an extreme point subset $\EX'(C) \subsetneq \EX(C)$ is obtained and $\frac{\sum_{\rv_C\in\EX'(C)}\rv_C}{|\EX'(C)|}$ is an approximation of $\rvS_C$. Then, $\oplus_{C \in \Pat^*}\frac{\sum_{\rv_C\in\EX'(C)}\rv_C}{|\EX'(C)|}$, the approximation of $\rvS_V$, can be obtained in time quadratic in the maximal subgame size $|\hat{C}|$.

\section{Conclusion}
\label{sec:Conclusion}

The coalition game model for the multiterminal data compression problem was formulated in term of the characteristic cost function, being the entropy function. The convexity of the game and the nonemptiness of the core followed from the submodularity of the entropy function. For a decomposable game, the direct sum of the Shapley values in the subgames constituted the Shapley value in the entire system. In this case, the complexity is essentially determined by the subgame of the maximum size and the finest decomposer determines the least computational complexity. The study showed that the finest minimizer can be determined in quadratic time and the reduction in complexity by the decomposition method for obtaining the Shapley value is large when the number of users increases. The incentive for the users to cooperate was also provided in this paper: The least source coding rates are attained in the multiterminal data compression problem only if the users cooperatively encode the multiple random sources.
One extension of the work in this paper is the design of the source coding scheme for the Shapley value: determining the codeword for source $\RZ{i}$ at rate $\rS_i$ for all $i \in V$.

\appendices

\section{}
\label{app:background}

For the entropy function $H \colon 2^V \mapsto \RealP$, the dual set function $\Hdual$ is defined as \cite[Section 2.3]{Fujishige2005}
$$ \Hdual(X) = H(V) - H(V \setminus X) = H(X | V \setminus X) $$
and is supermodular.\footnote{$f$ is supermodular if $-f$ is submodular.}
The polyhedron $P(\Hdual,\geq)$ contains all SW constrained rate vectors with sum-rate $r(V) \geq H(V)$, while the base polyhedron $B(\Hdual,\geq)$ contains all SW constrained rate vectors with exact sum-rate $r(V) = H(V)$, i.e., the SW constraints in \cite{SW1973,Cover1975} are given in terms of the dual set function $\Hdual$. So is the coalition game model proposed in \cite{Madiman2008ISIT,Madiman2008}.\footnote{The authors in \cite{Madiman2008ISIT,Madiman2008} formulated a game model $\G{V}{\Hdual}$ directly based on the SW constraints, where the dual set function $\Hdual(X)$ quantifies the payoff incurred when the users cooperate in coalition $X$. The convexity of the game was shown based on the supermodularity of $\Hdual$.}
Based on the equivalence $B(H,\leq) = B(\Hdual,\geq)$ \cite[Lemma 2.4]{Fujishige2005}, the data compression problem is studied in terms of the entropy function $H$ in this paper.

\begin{figure}[tbp]
	\centering
    \scalebox{0.8}{

\begin{tikzpicture}

\begin{axis}[
width=4in,
height=2.2in,
xlabel= {\Large $r_1$},
xmin=0,
xmax=6,
xmajorgrids,
ylabel = {\Large $r_2$},
ymin=0,
ymax=8,
ymajorgrids,
legend style={at={(0.8,0.3)},anchor=north west,draw=black,fill=white,legend cell align=left}
]

\addplot[
color=black,
line width=3.0pt,
only marks,
mark=asterisk,
mark options={solid}] coordinates {
(2.5,4.5)
};
\addlegendentry{Shapley value $\rvS_V$};

\addplot[
color=red,
line width=3.0pt,
only marks,
mark=triangle,
mark options={solid,,rotate=180}] coordinates {
(4,3)
(1,6)
};
\addlegendentry{$\EX(V)$};

\addplot[color=blue] coordinates {
(0,6)
(1,6)
(4,3)
(4,0)
};
\addplot[area legend,solid,fill=blue!20,opacity=4.000000e-01]coordinates {
(0,0)
(0,6)
(1,6)
(4,3)
(4,0)
(0,0)
};
\node at (axis cs:2,2.8) {\textcolor{blue}{$P(H,\leq)$}};

\addplot[color=red] coordinates {
(6,3)
(4,3)
(1,6)
(1,8)
};
\addplot[area legend,solid,fill=red!20,opacity=4.000000e-01]coordinates {
(6,8)
(6,3)
(4,3)
(1,6)
(1,8)
(6,8)
};
\node at (axis cs:4,6.5) {\textcolor{red}{$P(H^{\#},\geq)$}};

\addplot[color=orange, line width = 2] coordinates {
(4,3)
(1,6)
};
\node at (axis cs:3.5,3.5) [pin={[pin distance = 5mm, pin edge=black,ultra thick]80:\textcolor{black}{$B(H,\leq)=B(H^{\#},\geq)$}}] {};

\end{axis}

\end{tikzpicture}

}
	\caption{The polyhedra $P(H,\leq)$ and $P(\Hdual,\geq)$ of the entropy function $H$ and its dual set function $\Hdual$, respectively. The base polyhedra $B(H,\leq)$ and $B(\Hdual,\geq)$ coincide. The Shapley value $\rvS_V$ is also called the symmetric rate vector in \cite{Pradhan2000,Sartipi2005}. }
	\label{fig:HDual}
\end{figure}
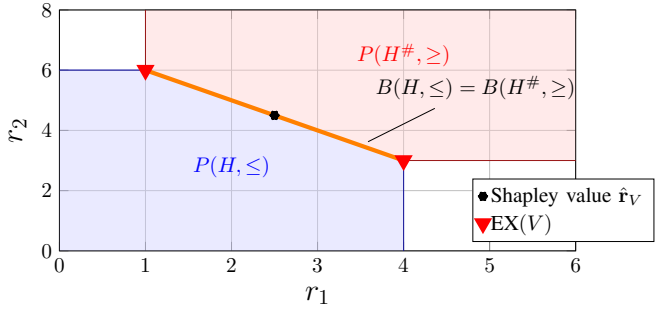

\begin{example} \label{ex:HDual}
    For a two terminal source with the entropy function: $H(\emptyset) = 0$, $H(\Set{1}) = 4$, $H(\Set{2}) = 6$ and $H(\Set{1,2}) = 7$. We plot polyhedra $P(H,\leq)$ and $P(\Hdual,\geq)$ in Fig.~\ref{fig:HDual}, where $B(H,\leq) = B(\Hdual,\geq)$ is the SW constrained achievable rate region $\RRCO(V)$.
\end{example}

Let $\Pi(V)$ be the set of all partitions of $V$. We have $H(V) = \min_{\Pat \in \Pi(V)} \sum_{C \in \Pat}H(C)$ \cite[Theorem 2.6]{Fujishige2005}. Each minimizer is a decomposer of the game $\G{V}{H}$ \cite[Theorems 3.36, 3.38 and 3.39]{Fujishige2005}\footnote{Decomposable convex game corresponds to the disconnected submodular system that is defined in \cite[Theorems 3.36, 3.38 and 3.39]{Fujishige2005}.} and all minimizers form a partition lattice \cite[Definition 3.9]{Narayanan1991PLP},\footnote{It is called the Dilworth truncation lattice in \cite[Definition 3.9]{Narayanan1991PLP} since the $\min_{\Pat \in \Pi(V)} \sum_{C \in \Pat}H(C)$ is called the Dilworth truncation of $H$.}
where the coarsest and finest partitions uniquely exist. This explains the uniqueness of the finest decomposer of a decomposable convex game in \cite{Shapley1971Convex}. It is for sure that $\Set{V}$ always belongs to this partition lattice. Based on Definition~\ref{def:DecomposeGame}, $\G{V}{H}$ is decomposable if the partition lattice contains partitions other than $\Set{V}$; otherwise, $\G{V}{H}$ is indecomposable.

\section{}
\label{app:mainproof}
    For a decomposer $\Pat \in \Pi'(V)$, we have $\EX(V) = \oplus_{C \in \Pat} \EX(C)$ due to the decomposition of the core $\Core{V} = \oplus_{C \in \Pat} \Core{C}$ in Lemma~\ref{lemma:MinSepProp}. Then,
    \begin{equation}
        \begin{aligned}
            \rvS_V & = \frac{\sum_{\rv_V \in \EX(V)} \rv_V }{|\EX(V)|}  \\
                   & = \frac{\sum_{\rv_V \in \oplus_{C \in \Pat} \EX(C)} \rv_V }{|\oplus_{C \in \Pat} \EX(C)|} \\
                   & = \frac{\oplus_{C \in \Pat} \Big( \prod_{C' \in \Pat \colon C' \neq C} |\EX(C')| \sum_{\rv_C \in \EX(C)} \rv_C \Big)}{ \prod_{C \in \Pat} |\EX(C)| } \\
                   & = \oplus_{C \in \Pat} \frac{\sum_{\rv_C \in \EX(C)} \rv_C}{|\EX(C)|}  \\
                   & = \oplus_{C \in \Pat} \rvS_C.
        \end{aligned}  \nonumber
    \end{equation}
Theorem~\ref{theo:DecompShapley} holds. \hfill\IEEEQED

\bibliographystyle{IEEEtran}
\bibliography{CGBIB}

\end{document}